\documentclass[12pt]{article}
\usepackage{graphicx}
\begin{document}
\centerline{\bf  Mixed algorithms in the Ising model} 
\centerline{\bf on directed Barab\'asi-Albert networks}
 
\bigskip
\centerline{F.W.S. Lima}
 
\bigskip
 Departamento de F\'{\i}sica, Universidade Federal do Piau\'{\i}, 
 
 57072-970 Teresina - PI, Brazil

\medskip
  e-mail: wel@ufpi.br
\bigskip
 
{\small Abstract: On directed Barab\'asi-Albert networks with two and
 seven  neighbours selected by each added site, the Ising model does not
 seem to show a spontaneous magnetisation. Instead, the decay time for 
 flipping of the magnetisation follows an Arrhenius law for Metropolis
 and Glauber algorithms, but for Wolff cluster flipping the
 magnetisation  decays exponentially with time. On these networks the
 magnetisation behaviour  of the Ising model, with Glauber, HeatBath,
 Metropolis, Wolf or Swendsen-Wang algorithm competing against Kawasaki
 dynamics, is studied by Monte Carlo simulations. We show that the model
 exhibits the phenomenon of self-organisation (= stationary equilibrium)
 defined in \cite{kawasaki} when
 Kawasaki dynamics is not dominant in its competition with Glauber, HeatBath
 and Swendsen-Wang algorithms. Only for Wolff cluster flipping the
 magnetisation, this phenomenon occurs after an exponentially decay of 
magnetisation with time. The Metropolis results are independent of competition. We also study the same  process of competition described above but with Kawasaki dynamics at the same temperature as the other algorithms. The obtained 
results are similar for Wolff cluster flipping, Metropolis and Swendsen-Wang 
algorithms but different for HeatBath.}
 
 
\bigskip

 {\bf Introduction}
 
 Sumour and Shabat \cite{sumour,sumourss} investigated Ising models on 
 directed Barab\'asi-Albert networks \cite{ba} with the usual Glauber
 dynamics.  No spontaneous magnetisation was 
 found, in contrast to the case of undirected  Barab\'asi-Albert networks
 \cite{alex,indekeu,bianconi} where a spontaneous magnetisation was
 found below a critical temperature which increases logarithmically with
 system size. More recently, Lima and Stauffer \cite{lima} simulated
 directed square, cubic and hypercubic lattices in two to five dimensions
 with heat bath dynamics in order to separate the network effects  form
 the effects of directedness. They also compared different spin flip
 algorithms, including cluster flips \cite{wang}, for
 Ising-Barab\'asi-Albert networks. They found a freezing-in of the 
 magnetisation similar to  \cite{sumour,sumourss}, following an Arrhenius
 law at least in low dimensions. This lack of a spontaneous magnetisation
 (in the usual sense)
 is consistent with the fact
 that if on a directed lattice a spin $S_j$ influences spin $S_i$, then
 spin $S_i$ in turn does not influence $S_j$, 
and there may be no well-defined total energy. Thus, they show that for
 the same  scale-free networks, different algorithms give different
 results. Now we study the self-organisation phenomenon in the Ising
 model on the directed Barab\'asi-Albert networks studied for
 \cite{lima}. We consider ferromagnetic Ising models, in which the system
 is in contact with a heat bath at temperature $T$ and is subject to an
 external flux of energy. These processes can be simulated by two
 competing dynamics: the contact with the heat bath is taken into account
 by the single spin-flip Glauber kinetics  and the flux of energy into
 the system is simulated by a process of the Kawasaki type
 \cite{kawasaki}, where we exchange nearest-neighbour spins, which
 preserves the order parameter of the model. In our case, we consider
 two dynamics Kawasaki type. The first is the dynamics Kawasaki at
 zero temperature, already mentioned above, where there are an exchange of
 spins that favors an
 increase in the energy of the
 system. This method \cite{kawasaki} means that continuously new energy is
 pumped into the system from an outside source. Therefore, this kind of 
 Kawasaki process is not the usual relaxational one, where Kawasaki dynamics
is in the same temperature the others algorithms that are competing. Here, we combine its with algorithms beyond Glauber: Metropolis,
 HeatBath, Swendsen-Wang and Single-Cluster Wollf algorithm. 
 
\bigskip
 
\begin{figure}[hbt]
\begin{center}
\includegraphics[angle=-90,scale=0.46]{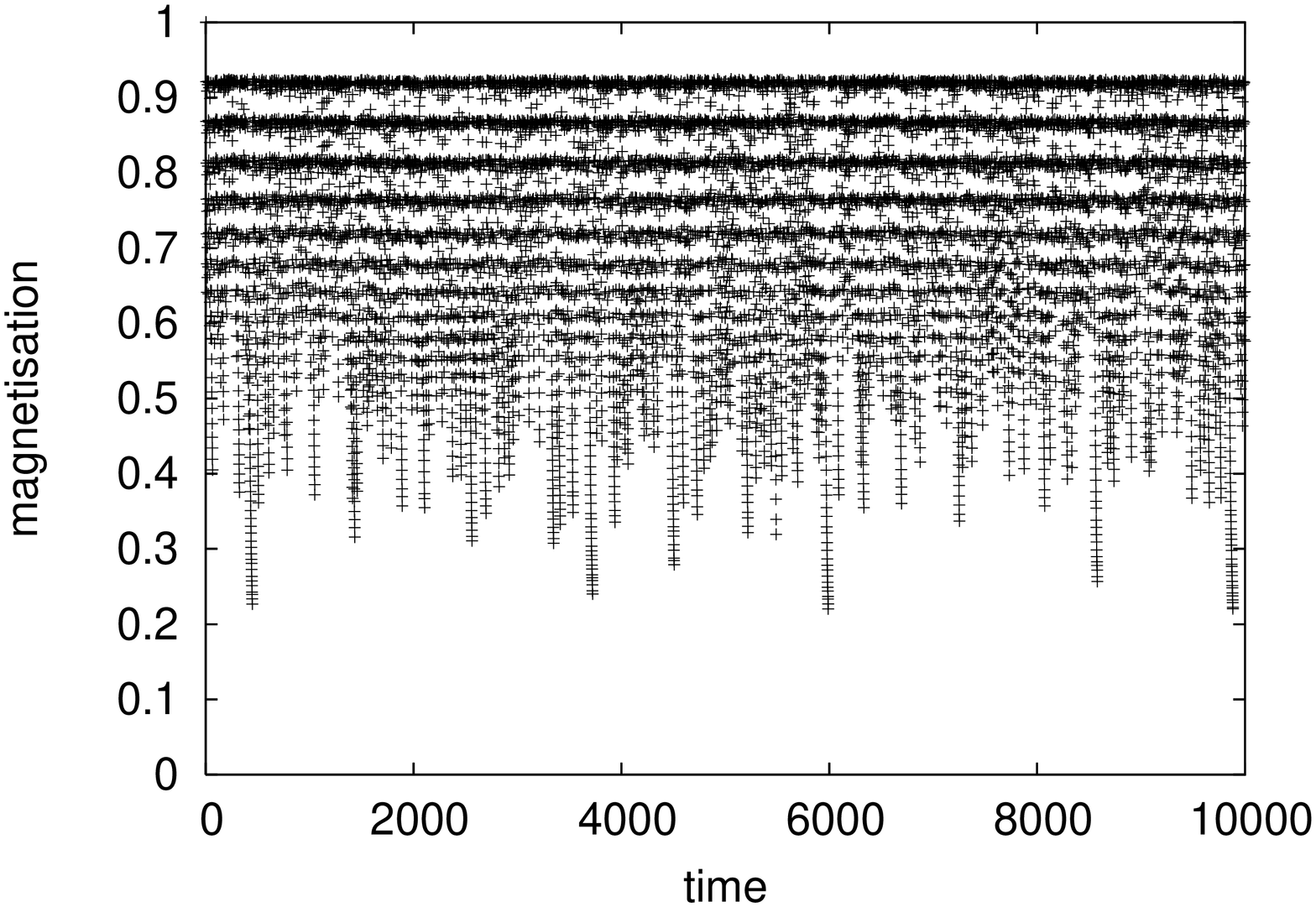}
\includegraphics[angle=-90,scale=0.46]{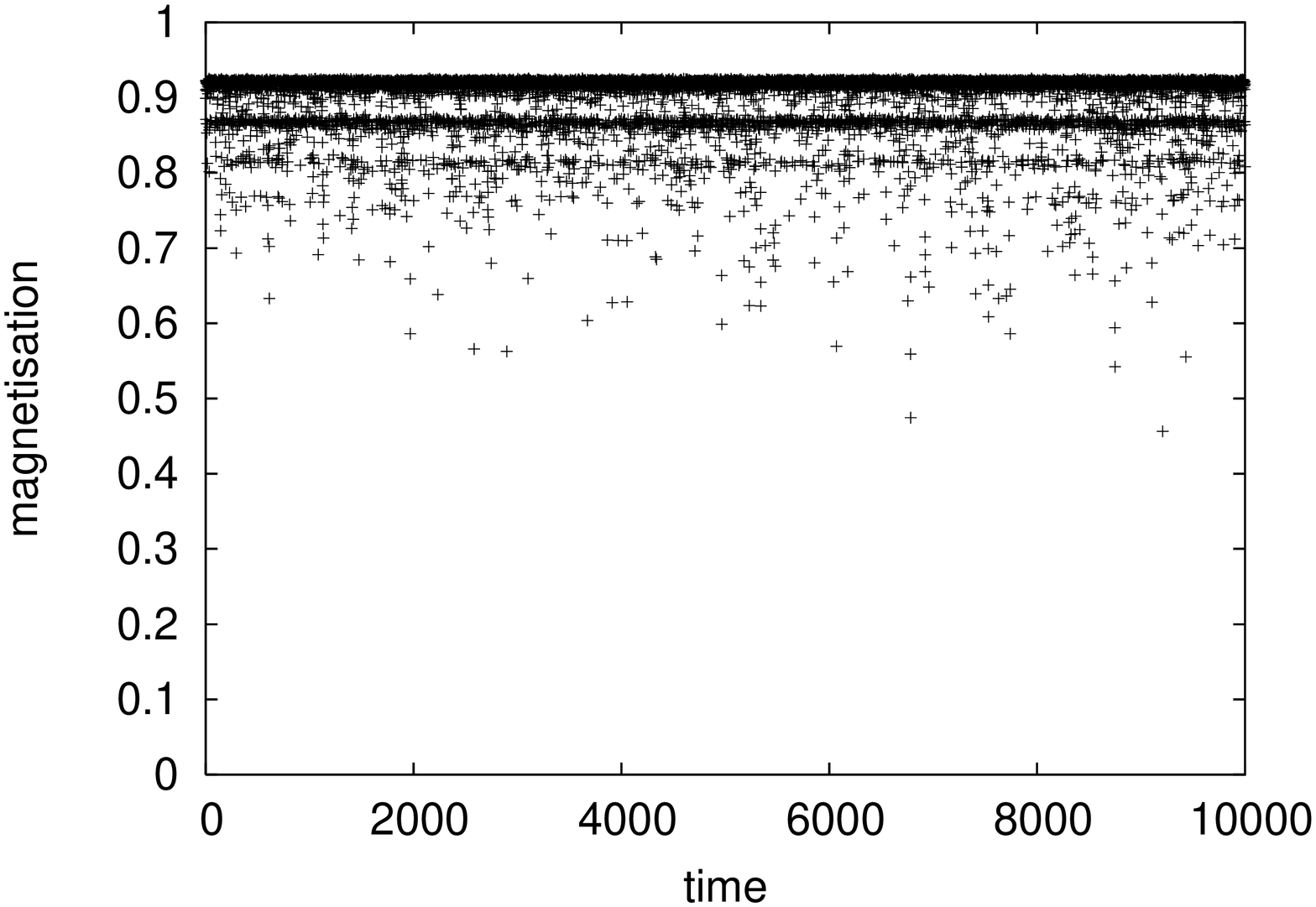}
\end{center}
\caption{
 Variation of normalised magnetisation on directed Barab\'asi-Albert
 network with half a million spins and $m=7$ at $k_BT/J = 1.7$, for
 HeatBath algorithm  competing against  Kawasaki dynamics with the zero  
temperature
at $p = 0.2$, that means for a predominance  of Kawasaki dynamics in Part a. Part b 
 for $p=0.8$, where predominance is of HeatBath algorithm.}
\end{figure}
  
\bigskip
 
\begin{figure}[hbt]
\begin{center}
\includegraphics[angle=-90,scale=0.46]{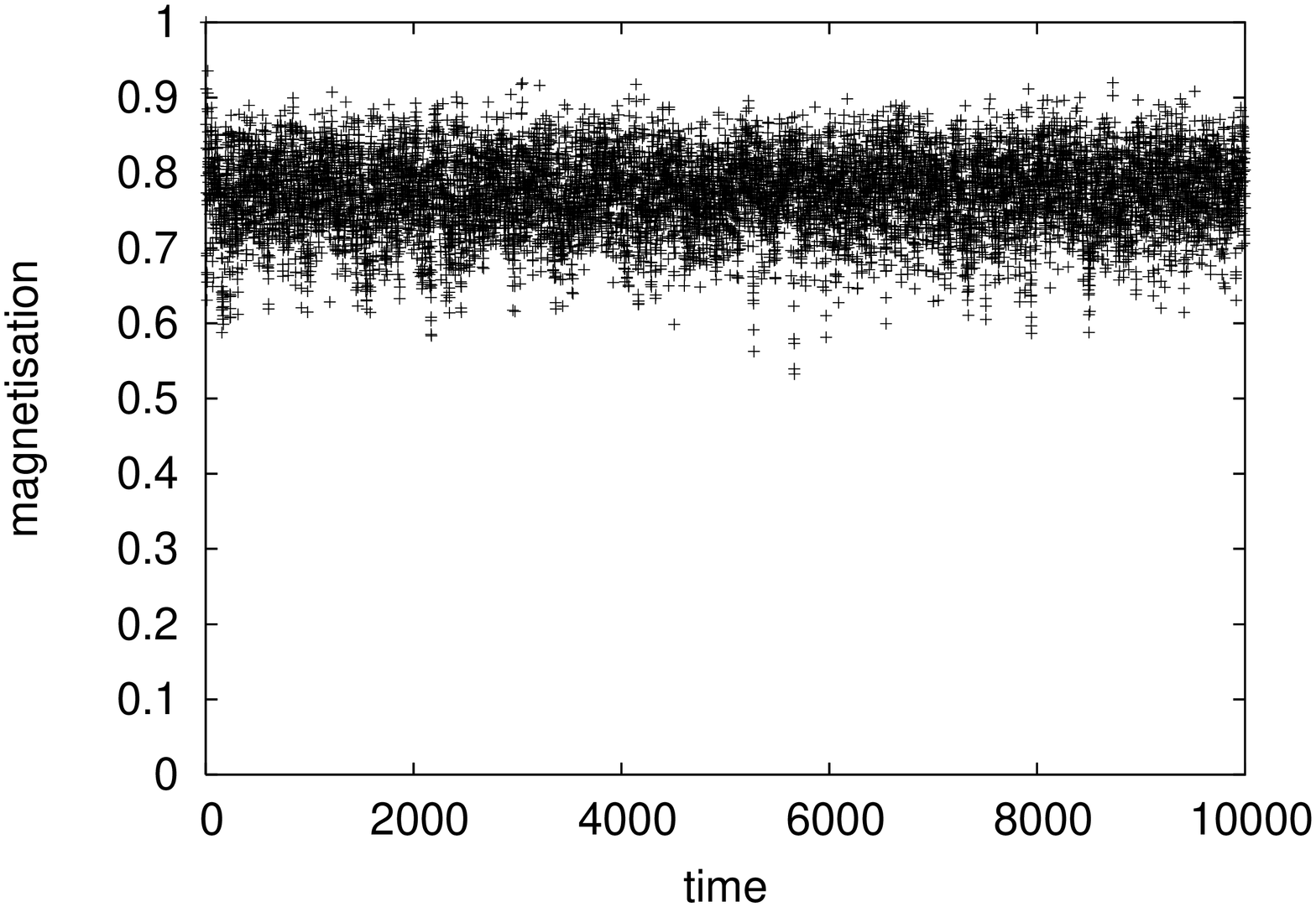}
\includegraphics[angle=-90,scale=0.46]{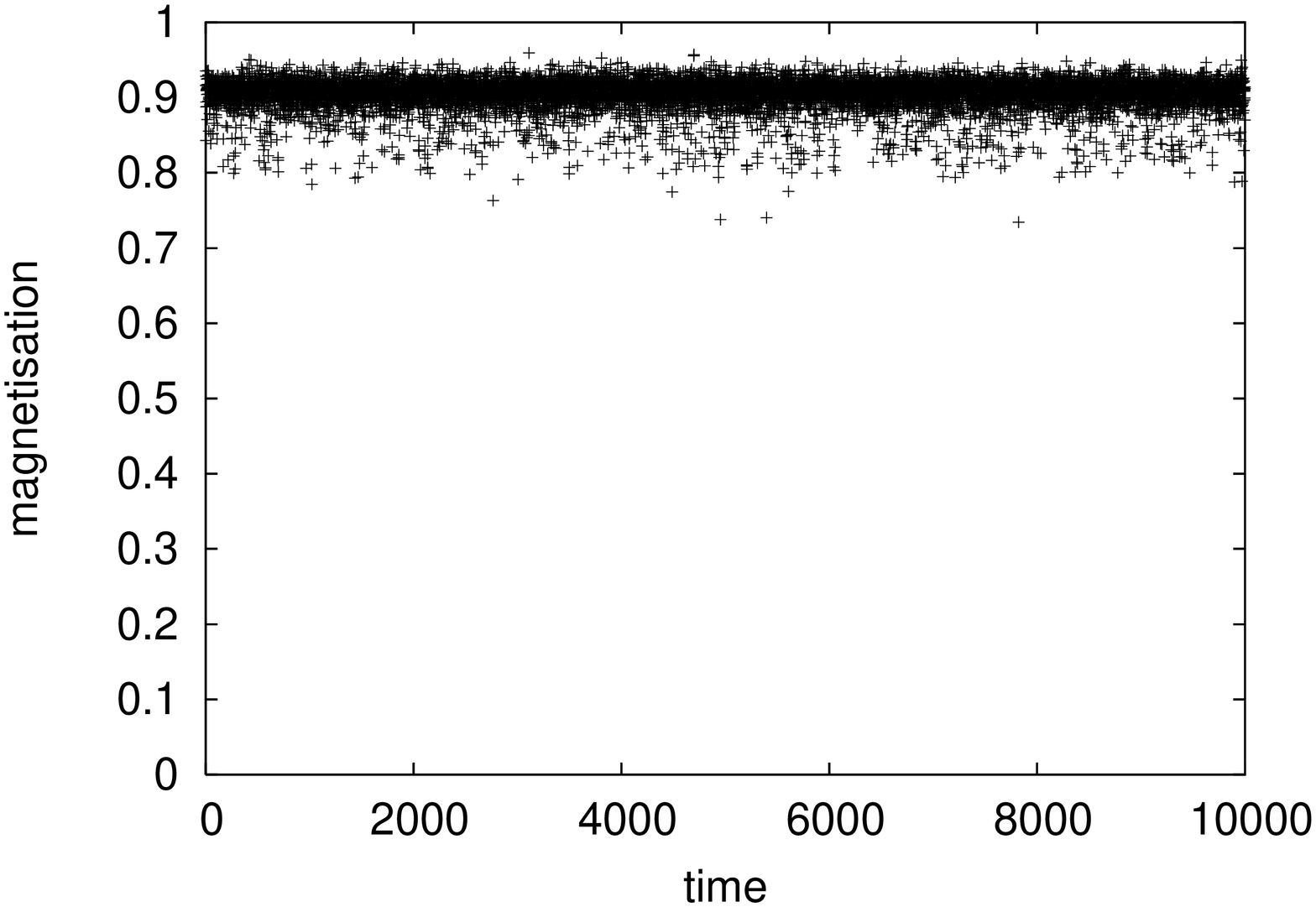}
\end{center}
\caption{
 As Fig.1 but Metropolis instead of HeatBath.}
\end{figure}
 
\begin{figure}[hbt]
\begin{center}
\includegraphics[angle=-90,scale=0.60]{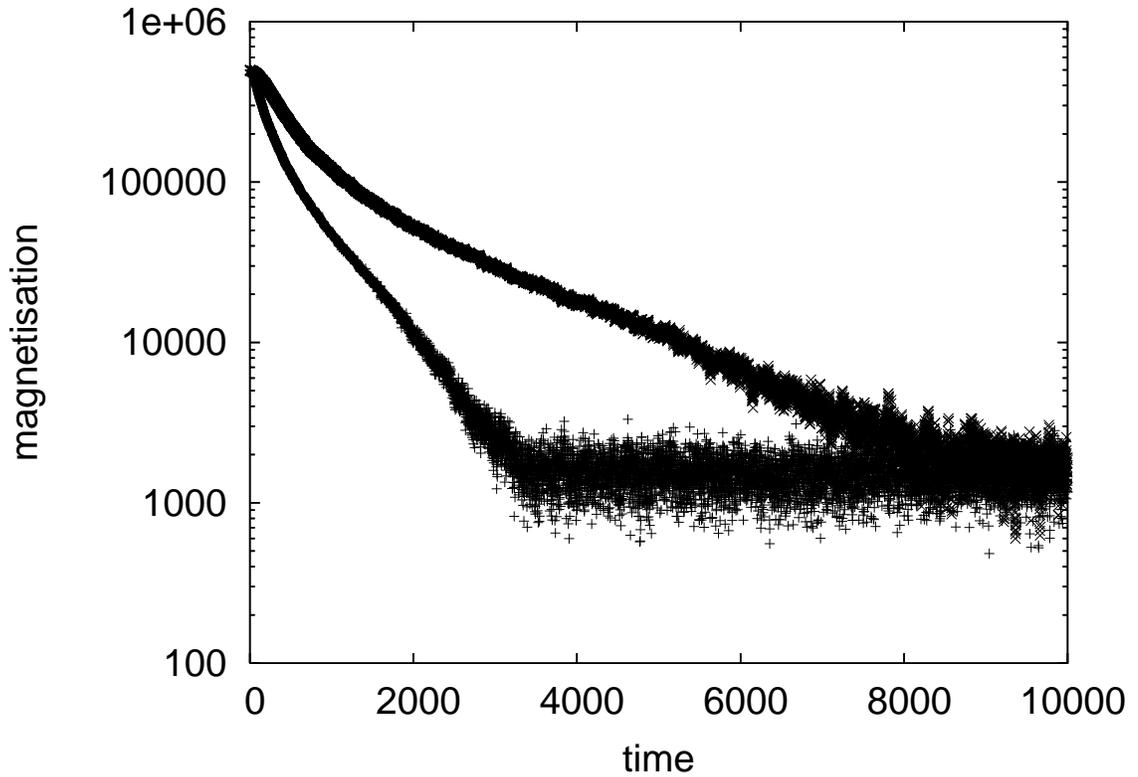}
\end{center}
\caption{Variation of magnetisation on directed Barab\'asi-Albert 
network with half a million spins and $m=7$ at $k_BT/J = 1.7, \;
p=0.2$ (+) and 0.8 (x) for competing Wolff algorithm and Kawasaki
dynamics with zero temperature. When there is a predominance of Kawasaki dynamics for
$p=0.2$, it makes systems  after an
exponential decay of normalised magnetisation self-organised \cite{kawasaki} already in shorter times in Wolff cluster flip algorithm.
} 
\end{figure}
 
\begin{figure}[hbt]
\begin{center}
\includegraphics[angle=-90,scale=0.60]{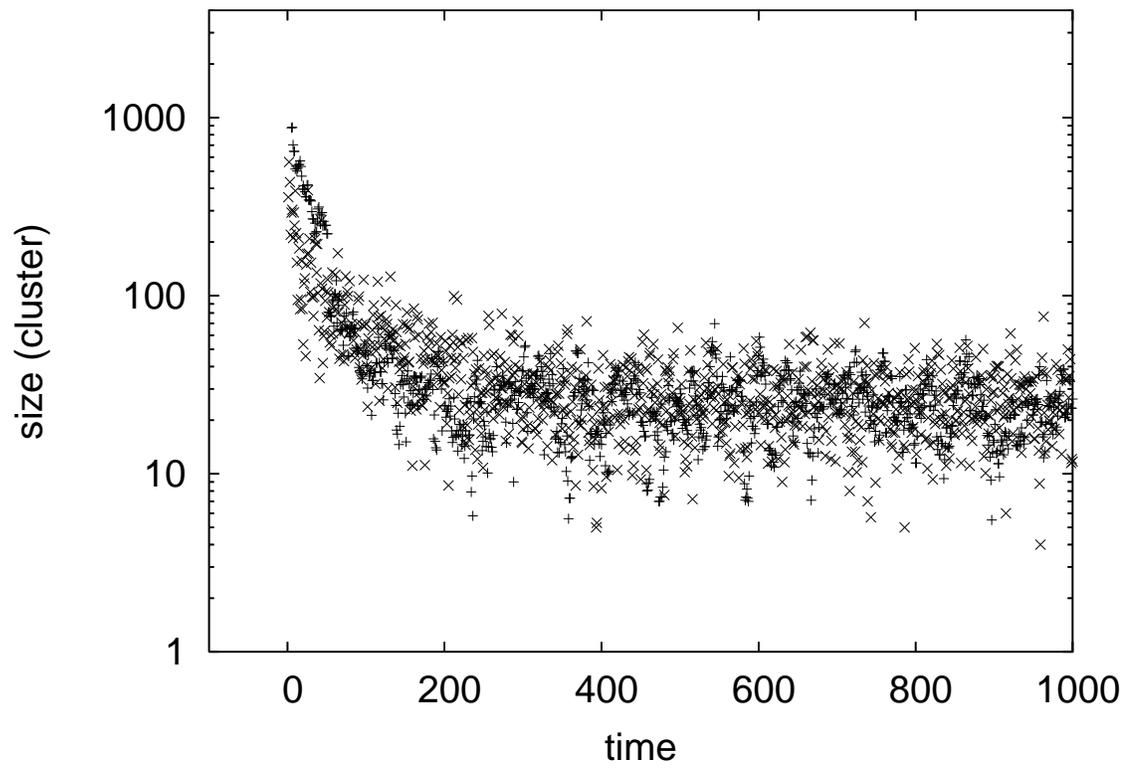}
\end{center}
\caption{
Cluster size for $p=0.2$(+) and $0.8$(x), with the zero temperature for
Kawasaki dynamics}
\end{figure}

\bigskip

\begin{figure}[hbt]
\begin{center}
\includegraphics[angle=-90,scale=0.46]{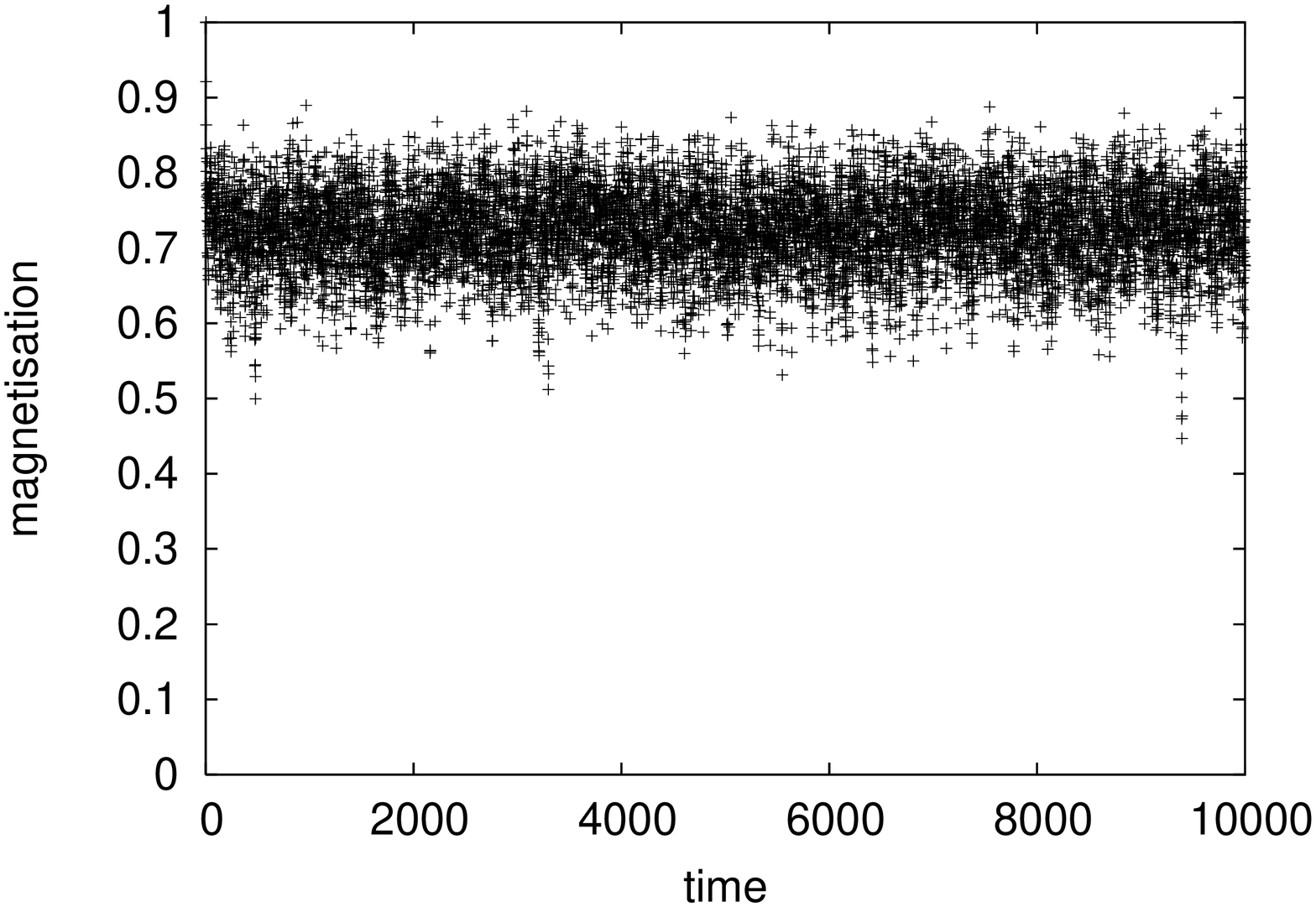}
\includegraphics[angle=-90,scale=0.46]{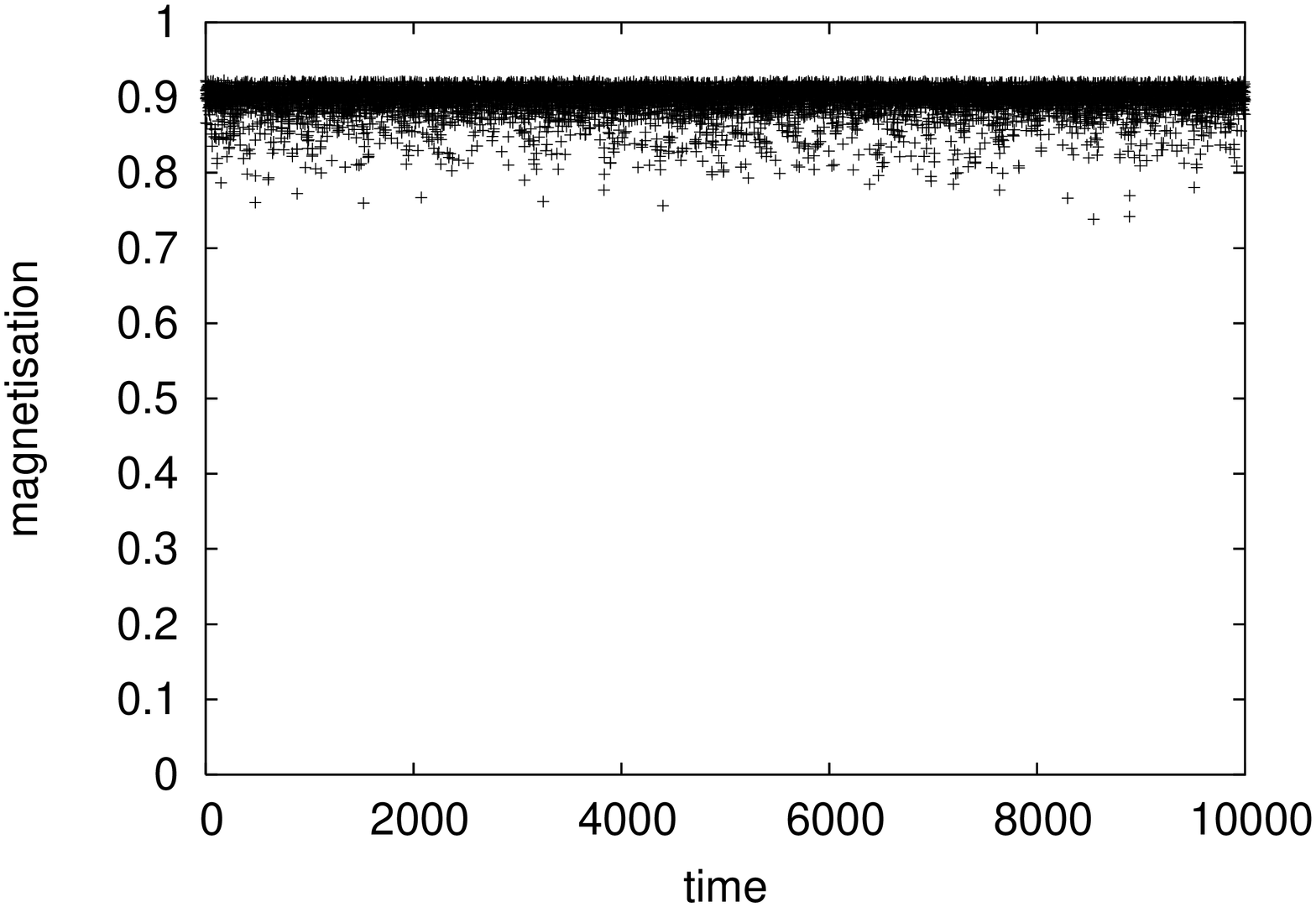}
\end{center}
\caption{
Variation of normalised magnetisation on directed Barab\'asi-Albert
 network with half a million spins and $m=7$ at $k_BT/J = 1.7$, for
 HeatBath algorithm  competing against Kawasaki dynamics with 
the same temperature as HeatBath algorithm at $p = 0.2$, that means for a 
predominance  of Kawasaki dynamics in Part a. Part b 
for $p=0.8$, where predominance is of HeatBath algorithm.}
\end{figure}
\bigskip
 
\begin{figure}[hbt]
\begin{center}
\includegraphics[angle=-90,scale=0.46]{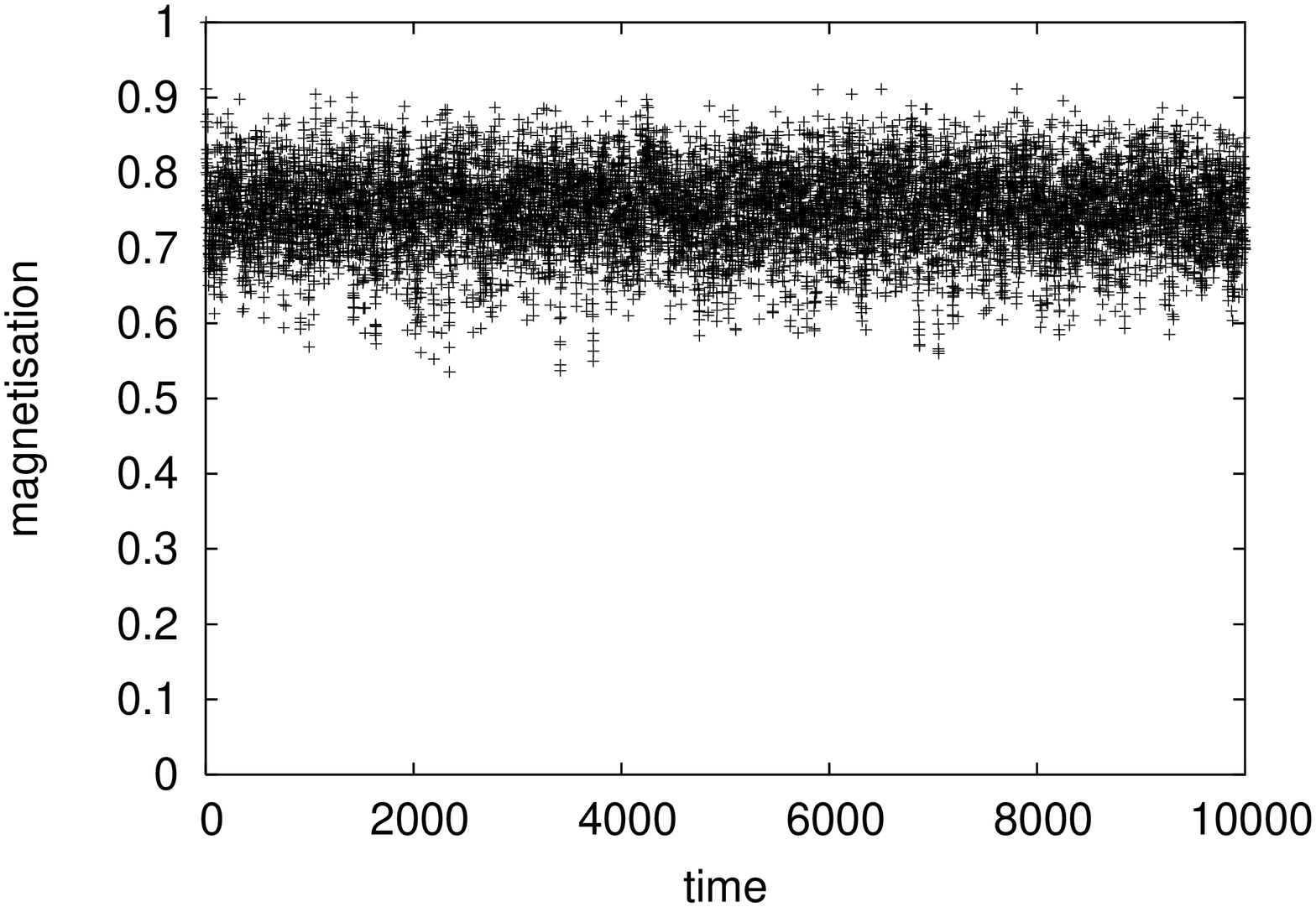}
\includegraphics[angle=-90,scale=0.46]{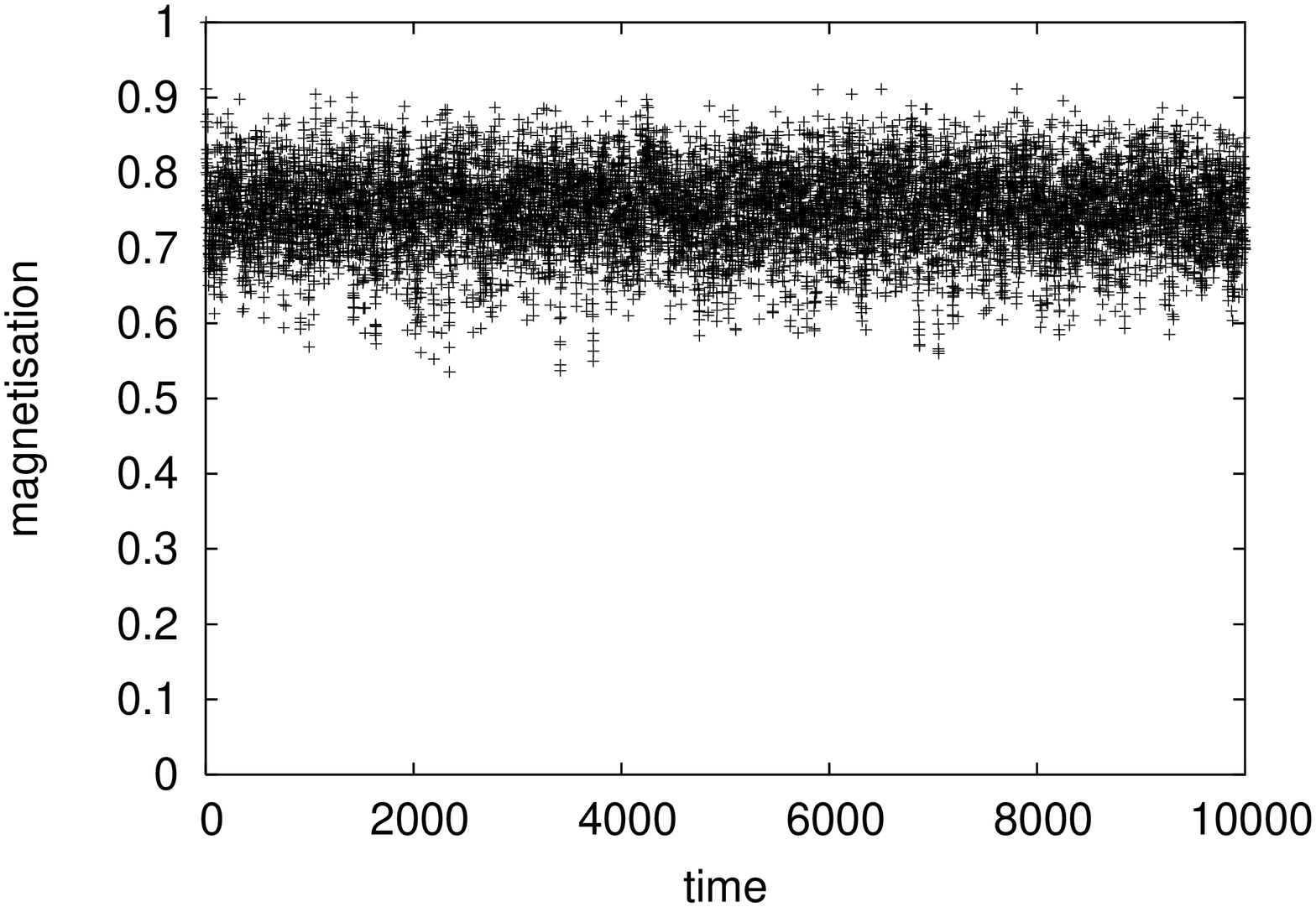}
\end{center}
\caption{
 As Fig.5 but Metropolis instead of HeatBath.}
\end{figure}

\begin{figure}[hbt]
\begin{center}
\includegraphics[angle=-90,scale=0.60]{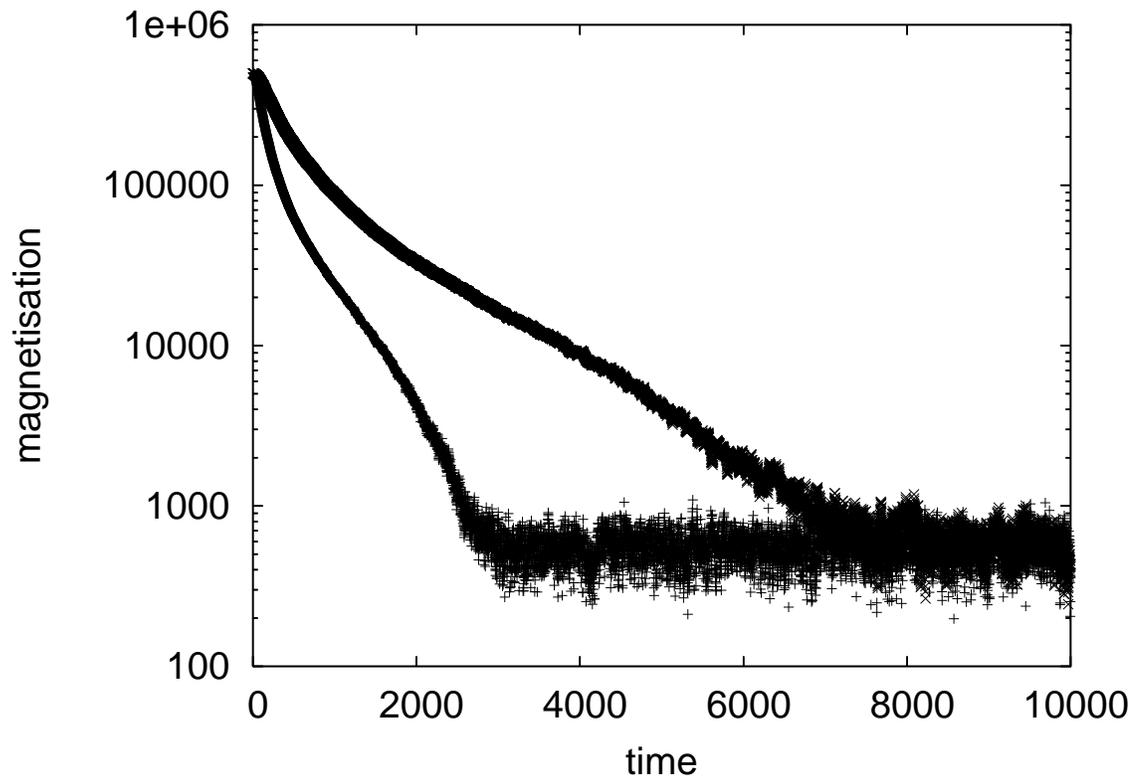}
\end{center}
\caption{Variation of magnetisation on directed Barab\'asi-Albert 
network with half a million spins and $m=7$ at $k_BT/J = 1.7, \;
p=0.2$ (+) and 0.8 (x) for competing Wolff algorithm and Kawasaki
dynamics with the same temperature for both.}
\end{figure}

\begin{figure}[hbt]
\begin{center}
\includegraphics[angle=-90,scale=0.60]{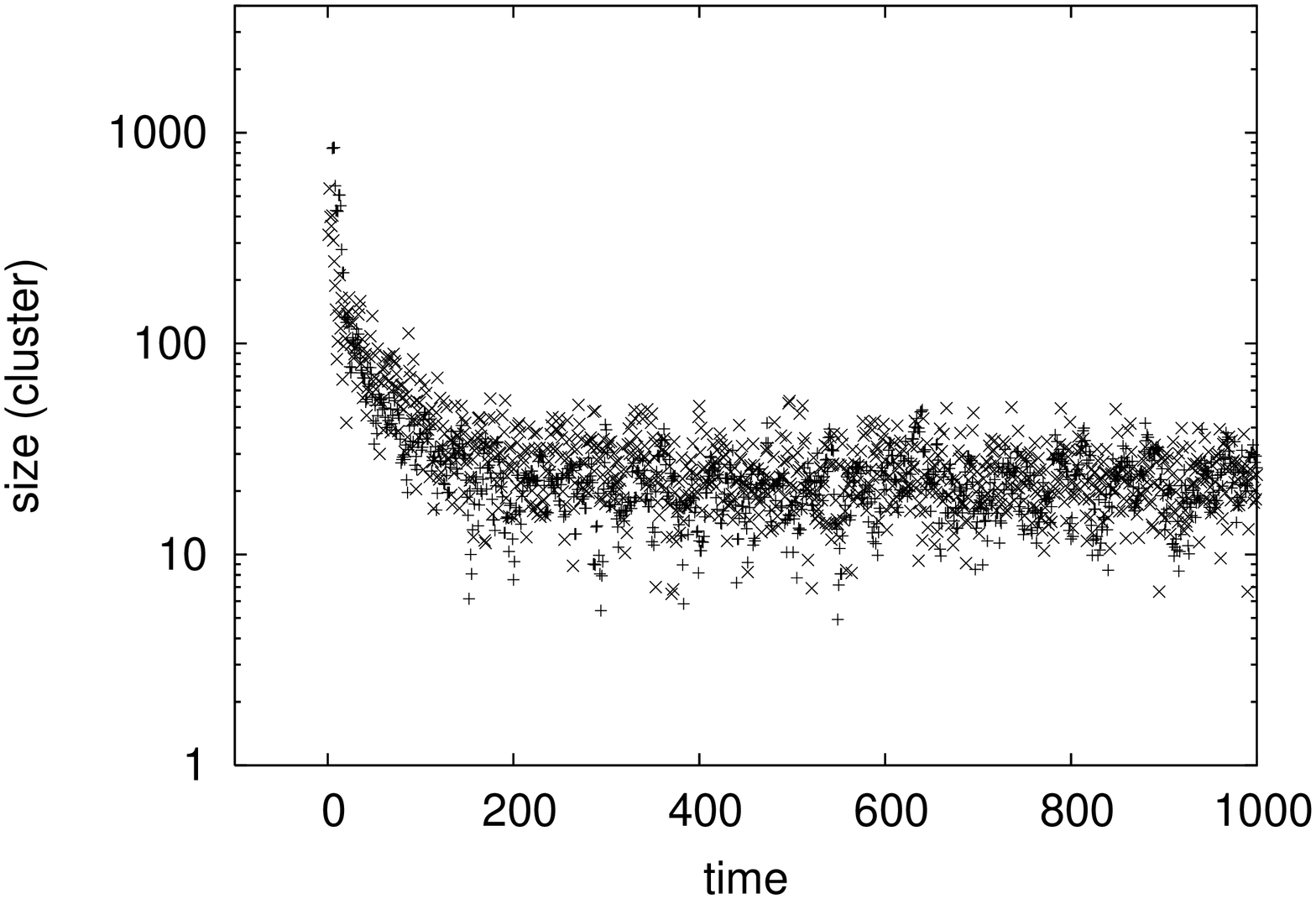}
\end{center}
\caption{Cluster size for $p=0.2$(+) and $0.8$(x) for Kawasaki dynamics with the same temperature as Wolff algorithm.} 
\end{figure}

\bigskip

{\bf Ising Model on Barab\'asi-Albert Networks}

We start with all spins up and always use half a 
million spins, with each new site added to the network selecting $m = 2$
 or 7 already existing sites as neighbours influencing it; the newly
 added spin does not influence these neighbours. The spin updates are
 made using with probability $p$ an algorithm that changes the order
parameter (= magnetisation), and with probability $q=1-p$  
 the Kawasaki dynamics which keeps the order parameter constant. The
 Metropolis transition probability of flipping spins is given by the
 well-known rate  $ w_{i}(\sigma) = \min[1,\exp(-\Delta E_{i}J/k_{B}T)]$,
 where $\Delta E_{i}$ is the energy change related to the given spin
 flip for local algorithms; we use the corresponding traditional probabilities
for global algorithms \cite{wang}. The others competing process, occuring
with a probability $1-p$, for two different Kawasaki dynamics studied here. The first one is the two-spin exchange Kawasaki dynamics at zero temperature
characterised by the transition probability of exchanging two
nearest-neighbour spins at sites $i$ and $j$:
 
\[ w_{ij}(\sigma) = \left\{
  \begin{array}{ll}
   0 , & \Delta E_{ij} \leq 0,\\
   1 , & \Delta E_{ij} > 0.
   \end{array}
   \right. 
\]
and 
 
\[ w_{i}(\sigma) = \left\{
  \begin{array}{ll}
   \exp^{(-\Delta E_{i}J/k_{B}T)} , & \Delta E_{i} > 0,\\
   1 , & {\rm otherwise}.
   \end{array}
   \right. 
\]
for Kawasaki dynamics in the same temperature the other algorithms.
 With $k_BT/J = 1.0$ and 1.7 and $p=1.0$, we confirmed
\cite{sumour,sumourss} the unusual  behaviour of the magnetisation,
 which for $m=7$ stays close to 1 for a long time inspite of
 fluctuations, 
 for HeatBath algorithm, Fig.1a. For $m=7$, $k_BT/J = 1.7$,
 and $p=0.2$  that means for a predominance of Kawasaki dynamics, 
 much smaller fluctuations occur around some magnetisation values 
(from top to bottom), after
 that, a big fluctuation occurs to a lower value of this magnetisation. This
 phenomenon occurs, because the  big energy flux through the Kawasaki 
 dynamics tries to self-organise the system \cite{kawasaki}, but with a small
 probability
 $p=0.2$ competes with the algorithm of HeatBath algorithm.  
 In  Fig.1b, where $p=0.8$ instead, the HeatBath algorithm is
 predominant, and the fluctuations occur only near two well defined 
 values of magnetisation, followed by a fall of magnetisation. In Fig.2a
 we have competing Metropolis algorithm and Kawasaki dynamics for
 $p=0.2$; this is similar to $p=0.8$, Fig.2b; the only difference between
 them is a reduction in the fluctuations, when the Metropolis algorithm is
 predominant. In Fig.3 we have competing Wolf and Kawasaki dynamics,
 where we see that the predominance of Kawasaki dynamics speeds up the
 system towards a self-organised state 
 (=stationary equilibrium \cite{kawasaki})
after that an exponential decay towards a value different from zero. For
 Swendsen-Wang cluster flips, for both $p=0.2$ and $p=0.8 $, the
 magnetisation scattered about zero (not shown). Only for competing Wolff
 cluster flips and Kawasaki dynamics, a nice  exponential decay towards
 different of zero value is found in Fig.3. In Fig.4 we observe
 that for $ p=0.2$ the formed cluster is bigger than for $p=0.8$. This
 explains the behavior of magnetisation to fall faster towards a
 stationary equilibrium when Kawasaki dynamics is predominant. 
In Fig.5, we observe that the magnetisation behavior for Kawasaki dynamics with temperature different from zero competing with algorithm HeatBath is 
different from Kawasaki dynamics at zero temperature (Fig.1) and insensitive to the value of the competition probability $p$. In Fig.6, for Metropolis algorithm
competing with Kawasaki dynamics with temperature different of zero, the magnetisation behavior is insensitive to the value of the competition probability $p$ as it occurs in Fig.5 and is identical to the behavior of Fig.2. In Fig.7 for Single-Cluster Wolff algorithm and Kawasaki dynamics competing at same temperature,
the magnetisation  behavior is as in Fig.3 for Kawasaki dynamics at zero temperature; the same similarity occurs with sizes of clusters: Fig.8 looks like Fig.4 despite the Kawasaki dynamics being different.

\bigskip
 
{\bf Conclusion}
 
In conclusion, we have presented a very simple nonequilibrium model on
directed Barab\'asi-Albert network \cite{sumour,sumourss}. We think that
this model, which exhibits the self-organisation phenomenon \cite{kawasaki}, due to the
competition between the  algorithms studied here and the configuration
dependent Kawasaki dynamics. Thus, even for the
same  scale-free networks, different algorithms competing with the
the zero temperature and temperature same the others algorithms Kawasaki dynamics  give different results. 

 It is a  pleasure to thank D. Stauffer for many suggestions and fruitful
 discussions during the development this work and  also for the revision
 of this paper. I also acknowledge the Brazilian agency FAPEPI
 (Teresina-Piau\'{\i}-Brasil) for  its financial support.

\end{document}